\documentclass[fleqn,usenatbib]{mnras}

\usepackage[T1]{fontenc}
\usepackage{ae,aecompl}

\usepackage{graphicx}	% Including figure files
\usepackage{amsmath}	% Advanced maths commands
\usepackage{amssymb}	% Extra maths symbols
\usepackage{lipsum}

\newcommand{\etacar}{$\eta$ Car} % Eta Carinae short name
\newcommand{\kmpers}{\,\rm{\,km \, s^{-1}}} % km/s
\newcommand{\masslossrate}{\,\rm{M_{\odot} yr^{-1}}} % mass-loss rate
\newcommand{\solarm}{\,\rm{M_{\odot}}} % solar mass
\newcommand{\solarl}{\,\rm{L_{\odot}}} % solar luminosity

\title[Tracing $\eta$ Carinae's colliding winds]{Tracing the colliding winds of $\eta$ Carinae in He I}

\author[D. Grant et al.]{
David Grant$^{1, 2}$\thanks{E-mail: david.grant@bristol.ac.uk},
Katherine Blundell$^{2}$\thanks{E-mail: katherine.blundell@physics.ox.ac.uk},
Emma Godden$^{2}$,
Steven Lee$^{3,4}$,\newauthor{}
and Chris McCowage$^{3,4}$
\\
$^{1}$University of Bristol, HH Wills Physics Laboratory, Tyndall Avenue, Bristol, BS8 1TL, UK\\
$^{2}$University of Oxford, Department of Physics, Keble Road, Oxford, OX1 3RH, UK\\
$^{3}$Anglo-Australian Telescope, Coonabarabran NSW 2357, Australia\\
$^{4}$Research School of Astronomy and Astrophysics, Australian National University, Canberra, ACT 2611, Australia\\
}

\date{Accepted xxxx. Received yyyy; in original form zzzz}

\pubyear{2022}

\begin{document}
\label{firstpage}
\pagerange{\pageref{firstpage}--\pageref{lastpage}}
\maketitle

\begin{abstract}
$\eta$ Carinae is an extremely luminous and energetic colliding-wind binary. The combination of its orbit and orientation, with respect to our line of sight, enables direct investigation of the conditions and geometry of the colliding winds. We analyse optical He I 5876 and 7065 \mbox{\normalfont\AA} line profiles from the Global Jet Watch observatories covering the last 1.3 orbital periods. The sustained coverage throughout apastron reveals the distinct dynamics of the emitting versus absorbing components: the emission lines follow orbital velocities whilst one of the absorption lines is detected only around apastron ($0.08 < \phi < 0.95$) and exhibits velocities that deviate substantially from the orbital motion. To interpret these deviations, we conjecture that this He I absorption component is formed in the post-shock primary wind, and is only detected when our line of sight intersects with the shock cone formed by the collision of the two winds. We formulate a geometrical model for the colliding winds in terms of a hyperboloid in which the opening angle and location of its apex are parameterised in terms of the ratio of the wind momentum of the primary star to that of companion. We fit this geometrical model to the absorption velocities, finding results that are concordant with the panchromatic observations and simulations of $\eta$ Carinae. The model presented here is an extremely sensitive probe of the exact geometry of the wind momentum balance of binary stars, and can be extended to probe the latitudinal dependence of mass loss.
\end{abstract}


%%%%%%%%%%%%%%%%%%%%%%%%%%%%%%%%%%%%%%%%%%%%%%%%%%

%%%%%%%%%%%%%%%%% BODY OF PAPER %%%%%%%%%%%%%%%%%%

\section{Introduction}
\label{sec:tracing_introduction}
% Todo: update software repo to ck-motion, with added tracing orbit motion.

Colliding-wind binaries are characterised by two massive stars both driving powerful stellar winds. These winds collide forming shock fronts in a cone-like shape at the surface where their wind momenta balance. In these shocks the gas is compressed and heated, leading to the production of x-rays \citep[][]{Prilutskii1976X-raysBinaries, Cherepashchuk1976DetectabilityRays, Luo1990X-raysWinds}, non-thermal emission \citep[][]{Pollock1987NewStars, White1994ParticleWinds, Hamaguchi2018}, as well as dust \citep[][]{Williams2008DustBinaries}. The presence of the shock cone may also modulate the UV and optical line profile variability during the binary's orbit \citep[][]{Stevens1993CollidingVariability, Szostek2012TracingBinaries}. 

$\eta$ Carinae (hereafter \etacar{}) is one of the most luminous and energetic colliding-wind binaries in the Milky Way having a luminosity of $L=5 \times 10^6 \solarl$ \citep[][]{Davidson1997ETAENVIRONMENT} and an observed long-term brightening due to a vanishing natural coronagraph \citep[][]{Damineli2021SpectroscopicCarinae, Gull2022EtaEjecta, Pickett2022ChangesOcculter}. The primary star is a Luminous Blue Variable exhibiting one of the strongest winds on record: a mass-loss rate of ${\sim}8.5 \times 10^{-4} \masslossrate$ and terminal wind velocity of ${\sim}420 \kmpers$ \citep[][]{Hillier2001, Groh2012OnSpectra, Clementel20153DApastron}.

The companion star, despite having not been observed directly, is estimated to have a mass-loss rate and terminal wind velocity of ${\sim}$(1-2)$ \times 10^{-5} \masslossrate$ and ${\sim}3000 \kmpers$, respectively. These are deduced from modelling the x-ray emission from \etacar{}'s colliding winds \citep[][]{Pittard2002, Okazaki2008ModellingCollision, Parkin2011SpiralingCarinae, Russell2016ModellingCarinae}. Photoionisation modelling of the spectral variability of \etacar{}'s Weigelt blobs \citep[][]{Weigelt1986EtaInterferometry} indicated that the companion may be a hot O-star, with an effective temperature between 34,000 and 38,000 K \citep[][]{Verner2002TheCarinae, Verner2005TheD}. However, as noted by \citet[][]{Smith2018} the companion's wind parameters may be more consistent with those of a hydrogen-poor Wolf-Rayet (WR) star. \citet[][]{Smith2018} described a possible formation channel for this present-day companion via a past merger-in-a-triple scenario, and this has been shown to be tenable by the models of \citet[][]{Hirai2021SimulatingEruption}. 

Whether the companion is an O-star or a WR star, it is known to generate a significant flux of Helium ionising photons ($\lambda < 504$ \mbox{\normalfont\AA}, $h\nu > 24.6$ eV) which significantly affects the ionisation structure of the system. There has been a substantial effort to map the ionisation structure of Helium within the central 155 au of \etacar{}. This is because the Helium lines can be used to probe the conditions within the colliding wind interface, as has been utilised for other colliding-wind binaries such as V444 Cygni \citep[][]{Marchenko1994TheI} and WR 140 \citep[][]{Williams2021ConditionsProfile}. 

As for \etacar{}, there has been some speculation as to whether the He I lines are formed in the primary wind \citep[][]{Nielsen2007Interactions, Humphreys2008} or in the shock cone of the colliding winds \citep[][]{Damineli2008ACarinae}. More recently, the three-dimensional radiative transfer simulations of \citet[][]{Clementel20153DApastron, Clementel2015}, taking into account the ionisation structure of the primary wind and the effect of the companion's radiation, resulted in detailed Helium ionisation maps for \etacar{}. Since then, spectro-interferometric observations, with the infra-red K-band beam-combiner GRAVITY at the VLTI, have enabled milliarcsecond resolution observations of the He I 2s-2p (2.0587 $\mu$m) line transition \citep[][]{Sanchez-Bermudez2018} and these results show good agreement with the Helium ionisation maps of \citet[][]{Clementel20153DApastron}. In combination with these Helium ionisation maps, the estimated orbit \citep[][]{Damineli2000, Grant2020Uncovering140, Grant2021ProbabilisticBinaries} and orientation \citep[][]{Madura2012ConstrainingEmission, Teodoro2016} means that our line of sight intercepts the shock cone, providing an exceptional opportunity to study the conditions and geometry of \etacar{}'s colliding winds.

In this study, we investigate the geometry of \etacar{}'s colliding winds through time-series spectroscopic observations of the optical He I line profiles. In Section \ref{sec:tracing_observations}, we present observations from the Global Jet Watch's campaign on \etacar{}. We show that at different orbital phases the He I absorption varies, exhibiting distinct components at periastron and apastron. In Section \ref{sec:tracing_the_colliding_winds}, we employ multi-Gaussian fits to track the dynamics of these He I components, revealing that one of the absorption component's velocities deviates substantially from the expected orbital motion. We then describe a geometrical model of the colliding winds, cognisant of the known Helium ionisation maps, from which we simulate the He I absorption velocities. We fit this model to the Global Jet Watch data and tabulate the inferred parameters. In Section \ref{sec:tracing_discussion}, we discuss the latitudinal dependence and assumptions of this geometrical model. Finally, in Section \ref{sec:tracing_summary_and_conclusions} we summarise our findings.

\begin{figure*}
	\includegraphics[width=\textwidth]{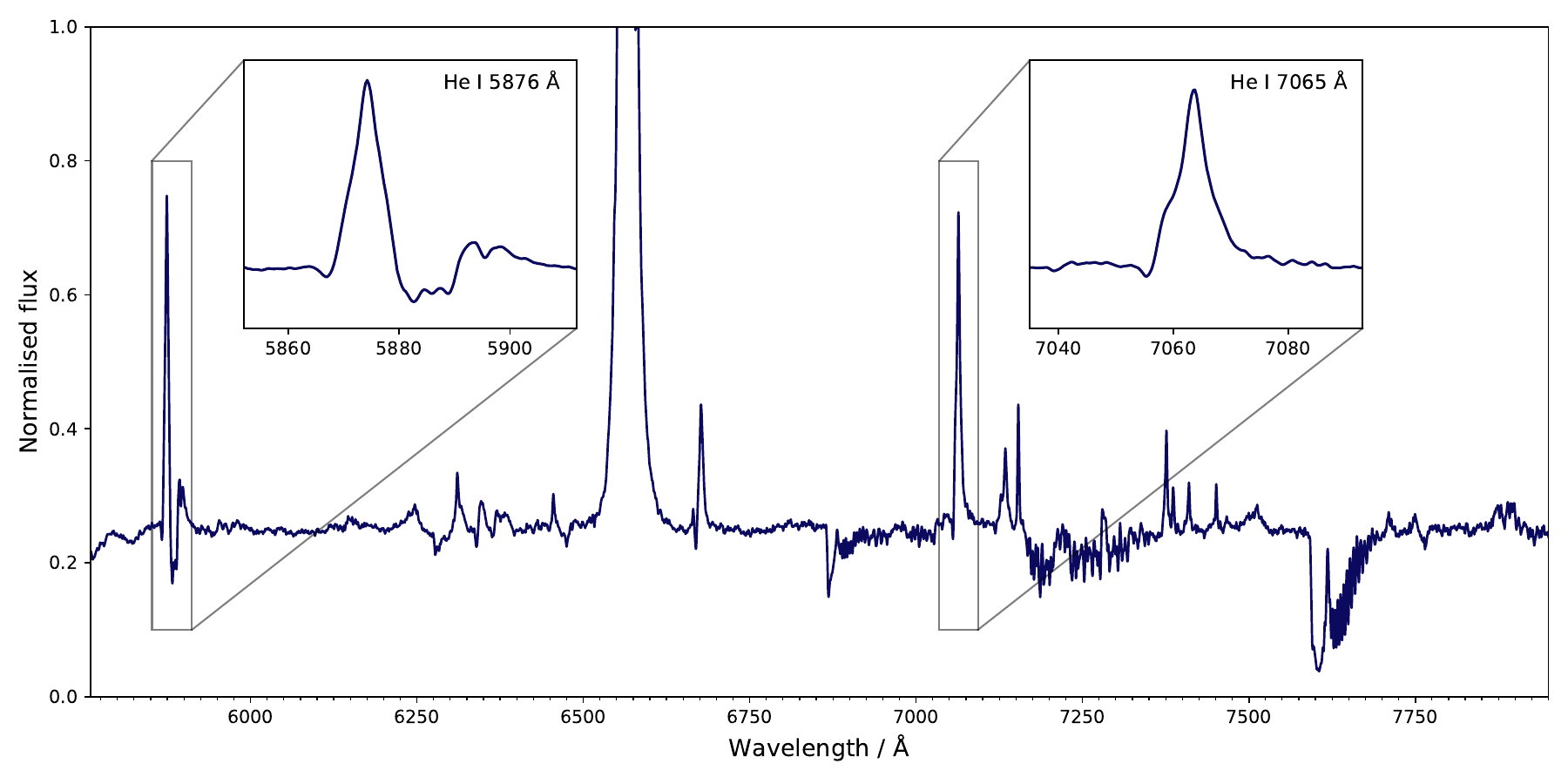}
    \caption{Example spectrum of \etacar{} taken from the Global Jet Watch observatory in Western Australia. Here we show the data prior to telluric correction but with the continuum removed. Inset panels display the two He I lines analysed in this work.}
    \label{fig:example_spec}
\end{figure*}

\begin{figure*}
	\includegraphics[width=\textwidth]{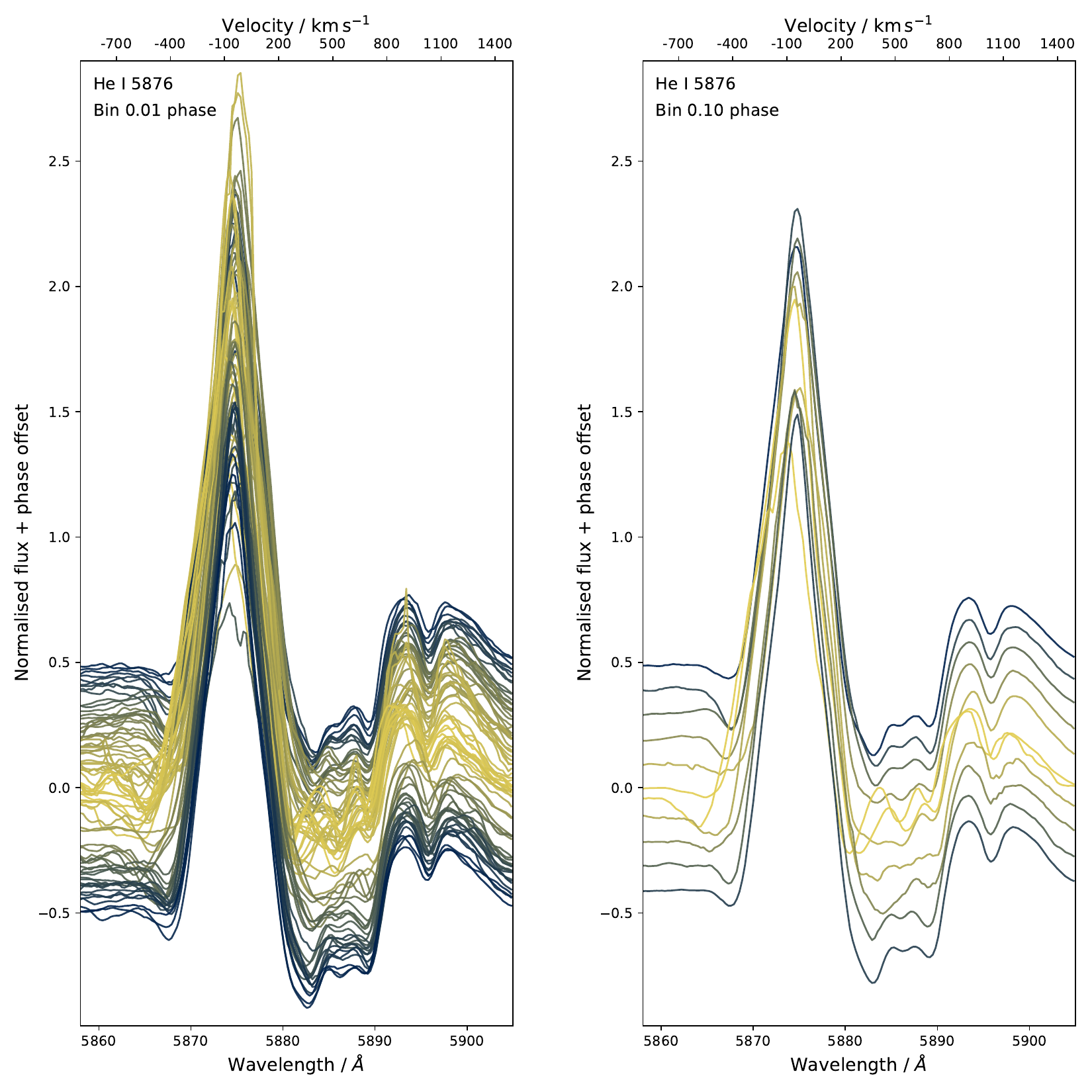}
    \caption{He I 5876 \mbox{\normalfont\AA} spectral line profiles binned into 0.01 phase bins (left) and 0.10 phase bins (right). The continuum has been removed and the y-offset corresponds to the orbital phase. The profiles are colour-coded with brighter yellow indicating the proximity to \etacar{}'s periastron. The corresponding velocities are shown on the top x-axes.}
    \label{fig:HeI5876_time_series_spectra}
\end{figure*}

\begin{figure*}
	\includegraphics[width=\textwidth]{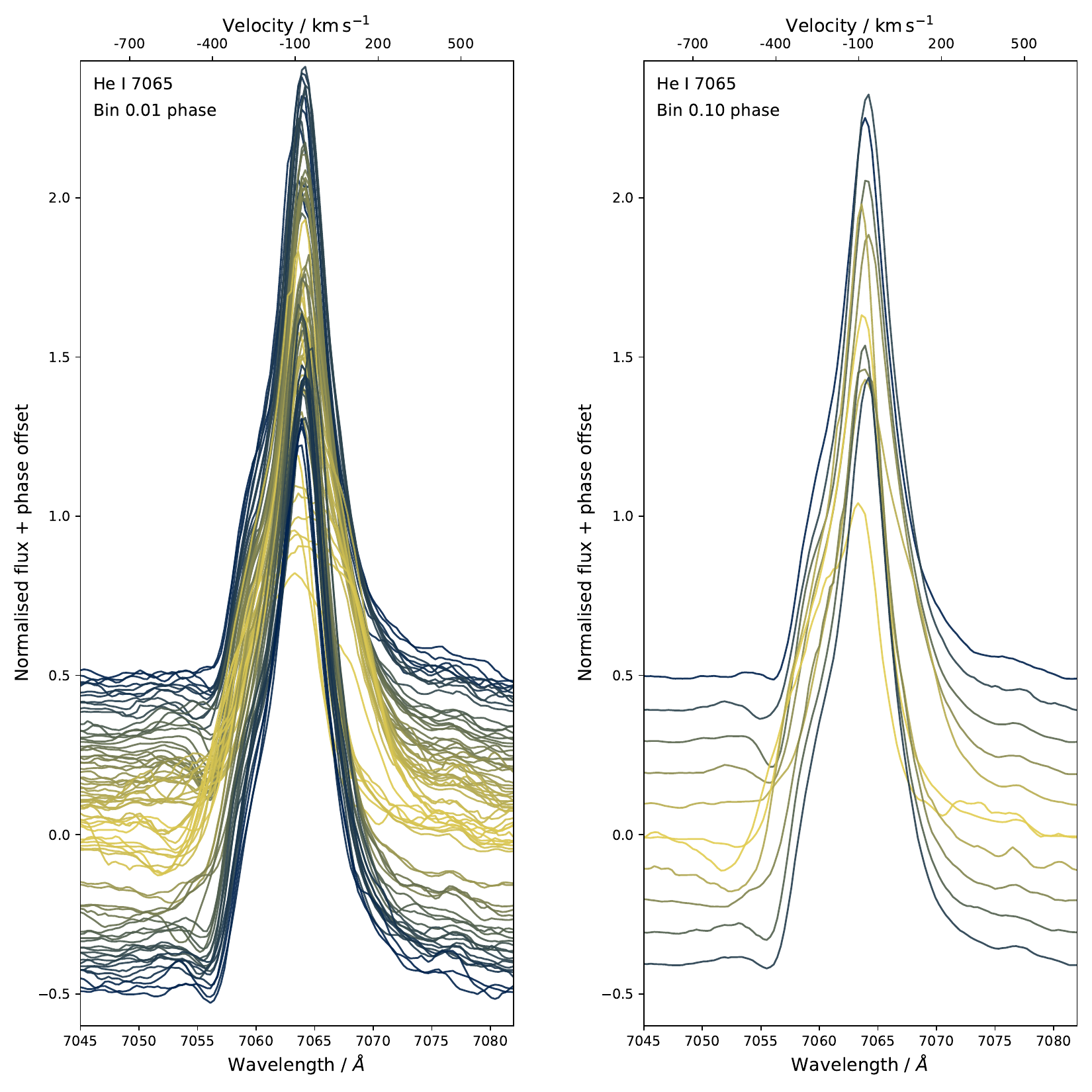}
    \caption{The same as Figure \ref{fig:HeI5876_time_series_spectra} but for the He I 7065 \mbox{\normalfont\AA} spectral line profiles.}
    \label{fig:HeI7065_time_series_spectra}
\end{figure*}

\section{Observations}
\label{sec:tracing_observations}
The Global Jet Watch is a network of five remotely controlled observatories separated in longitude to enable round-the-clock observing. The observatories are situated in South Africa, India, on the west and east coasts of Australia, and Chile. Each observatory has a fibre-fed spectrograph, and a 0.5 m telescope, that collects mid-resolution ($R \,{\sim} 4000$) optical data in the wavelength range ${\sim}5800-8400$ \mbox{\normalfont\AA}. The spectra are reduced by a bespoke data-reduction pipeline which includes dark correction, flat-field correction, and wavelength calibration derived from Thorium-Krypton frames. Full details of the observatories and instruments will be described in Blundell et al. in prep and Lee et al. in prep, respectively.

The spectra have their telluric absorption corrected for using \texttt{TelFit} \citep{Gullikson_2014}. This software is a wrapper to the line-by-line radiative transfer model \citep[LBLRTM][]{Clough_2005}. LBLRTM is a well tested and reliable radiative transfer model that is both accurate and efficient for modelling telluric absorption. For each spectrum, \texttt{TelFit} is used to produce a model of the tellurics by fitting the pressure, temperature, humidity, and resolution to the spectral data. Other important observation parameters (latitude, altitude, zenith angle) are set from the observatory meta data. During the fitting process, \texttt{TelFit} is given specific wavelength regions to ignore from the data which correspond to strong emission lines in the data. The resulting model telluric spectra are then subtracted from the \etacar{} spectra.

The Global Jet Watch observing campaign of \etacar{} began in early-2014 and since has amounted 5858 spectra across 1.3 orbital periods (2630 days), spanning the 2014 and 2020 periastra. These observations have exposure times ranging from 1 second to 300 seconds to capture both the H-alpha and He I lines at acceptable signal-to-noise ratios while avoiding saturation. Pertinent to this study, the observations are densely time-sampled throughout the orbital period, not just at periastron, and thereby provide excellent coverage of \etacar{}'s apastron. An example spectrum is displayed in Figure \ref{fig:example_spec}.

\subsection{The He I line profiles}
\label{subsec:he_i_line_profiles}

In this paper, we focus our analysis on the He I line profiles. As such, we select the 3554 Global Jet Watch spectra having exposure times of at least 100 seconds. These spectra are normalised by their local continuum and are barycentric corrected to the Solar System using the \texttt{barycorrpy} package \citep[][]{Kanodia2018Pythonbarycorrpy}. To further improve the signal-to-noise ratio, especially for the weaker blue-shifted absorption components, we median stack the spectra into orbital phase bins of width 0.01. These phases bins are computed according to the period, $P=2022.7$ d, of \citet[][]{Damineli2008TheEvents}, and the time of periastron, $T_0=2454850.1$ (JD) of \citet[][]{Grant2021ProbabilisticBinaries}. The He I lines have signal-to-noise ratios of ${\sim}$100 at the continuum level and of ${\sim}$200-300 at the peak of the lines.

\begin{figure*}
	\includegraphics[width=0.80\textwidth]{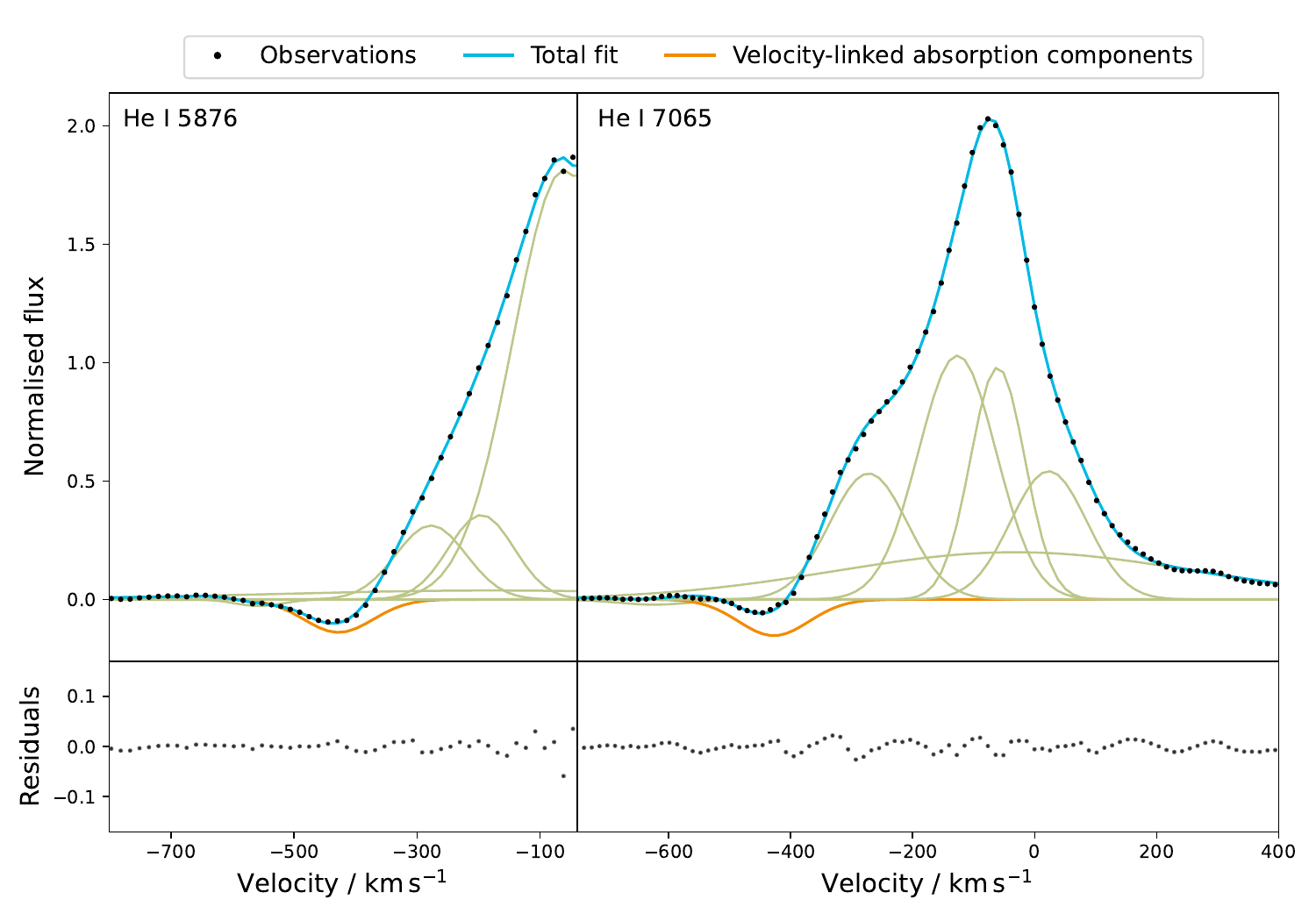}
    \caption{Example simultaneous multi-Gaussian fitting of the He I 5876 and 7065 \mbox{\normalfont\AA} spectral line profiles at $\phi = 0.61$. The orange absorption component has its velocity linked between the two lines. The light blue line is the total model. The bottom of each panel shows the residuals of each fit.}
    \label{fig:HeI7065_multiG_fits}
\end{figure*}

\begin{figure}
	\includegraphics[width=\columnwidth]{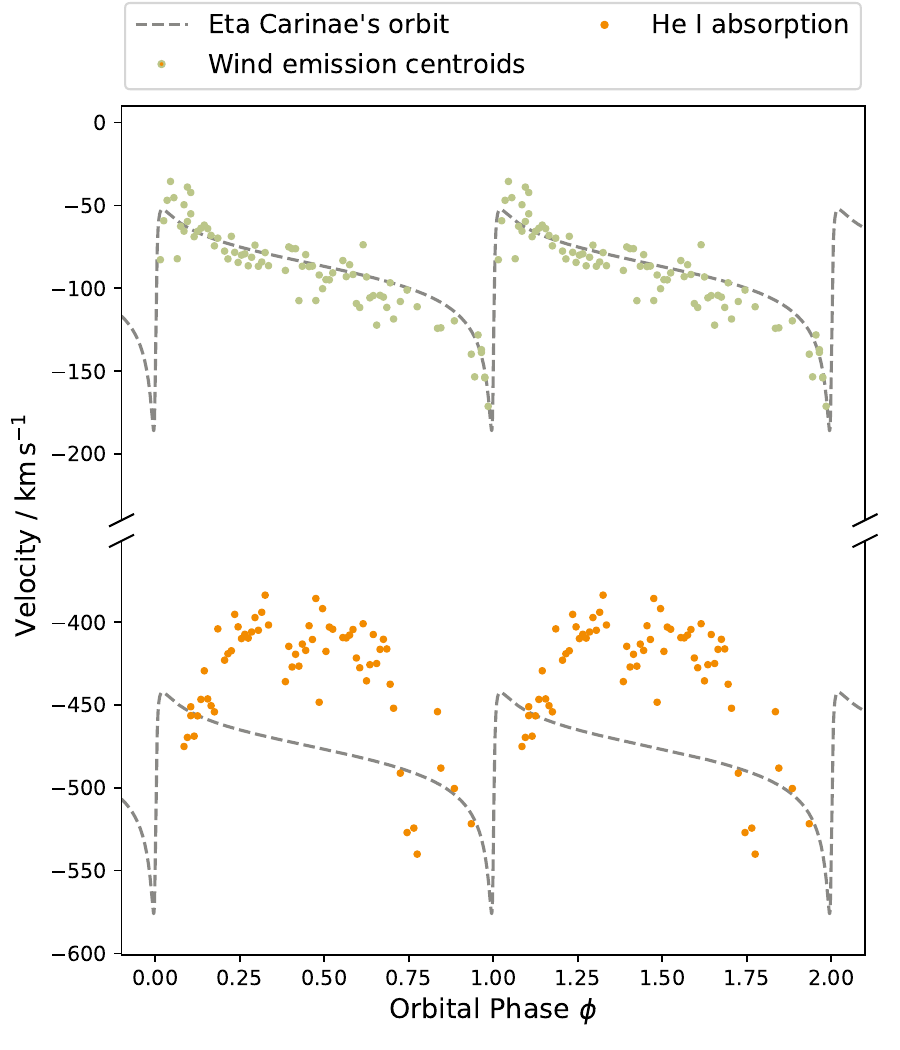}
    \caption{Emission (green) and absorption (orange) velocities extracted from the multi-Gaussian fits to the He I spectral lines. \etacar{}'s radial velocity curve estimated by \citet[][]{Grant2021ProbabilisticBinaries} is superimposed for comparison (dashed lines + offsets). The absorption represents the velocity-linked absorption component shown in orange in Figure \ref{fig:HeI7065_multiG_fits}.}
    \label{fig:HeI7065_dynamics}
\end{figure}

Initial analysis of the three He I lines in the Global Jet Watch spectra shows that both lines at 5876 \mbox{\normalfont\AA} and 6678 \mbox{\normalfont\AA} suffer blending from other species. As has been previously noted by \citet[][]{Richardson2015TheEvent}, the red wing of the He I 5876 \mbox{\normalfont\AA} line is blended with the complex Na I D doublet, and the blue wing of the He I 6678 \mbox{\normalfont\AA} line is blended with a [Ni II] line. As a result, we restrict our analysis to the unencumbered He I 7065 \mbox{\normalfont\AA} line, as well as the blue wing of the He I 5876 \mbox{\normalfont\AA} line. From these lines we can obtain clear dynamical information from the blue-shifted absorption components from both lines, and from the emission component of the He I 7065 \mbox{\normalfont\AA} line.

In Figures \ref{fig:HeI5876_time_series_spectra} and \ref{fig:HeI7065_time_series_spectra} we plot all of our 0.01 phase-binned spectra for the He I 5876 and 7065\mbox{\normalfont\AA} lines, respectively. Alongside these spectra we also present the data binned further into bins of 0.1 phase, to further elucidate the changing morphology of the emitting and absorbing components throughout \etacar{}'s orbit. In terms of emission we observe the expected variability: a low excitation event at periastron in which the narrow emission component disappears and the bulk velocity of the emission region shifts violently with the extremely eccentric periastron passage \citep[][]{Groh2004EtaEvent, Nielsen2007Interactions, Humphreys2008, Damineli2008TheEvents, Damineli2008ACarinae, Mehner2010, Mehner.A., Richardson2015TheEvent}. As for the blue-shifted absorption, we discern two distinct components in these line profiles. At periastron we observe strong absorption, at a velocity of ${\sim}{-500}\kmpers{}$, which has been detected and analysed previously \citep[e.g.,][]{Richardson2015TheEvent}. However, a second component, found at slower negative velocities, is detected only either side of periastron. In Figures \ref{fig:HeI5876_time_series_spectra} and \ref{fig:HeI7065_time_series_spectra} this behaviour is apparent, where the line profiles corresponding to phases around apastron clearly show this absorption component. We measure this second component to disappear at $\phi=0.95$ and reappear at $\phi=0.08$.

\subsection{He I absorption dynamics}
\label{subsec:he_i_absorption_dynamics}
To extract velocities from the He I 5876 and 7065 \mbox{\normalfont\AA} lines we employ multi-Gaussian fitting to decompose the complex line profiles into their constituent dynamical components. Following \citet[][]{Nielsen2007Interactions}, we construct a multi-Gaussian template comprised of four Gaussians to model the main emission, two negative Gaussians to model the blue-shifted absorption, and a low-amplitude broad Gaussian to model the extended wings of the profile. For the He I 5876 \mbox{\normalfont\AA} line we only use three Gaussians to model the emission, because, as noted in Section \ref{subsec:he_i_line_profiles}, we only consider the blue wing ($\lambda < 5875\,\rm{\mbox{\normalfont\AA}}$) due to the impact of Na absorption at longer wavelengths.

The Gaussian parameters are optimised using a Trust Region Reflective algorithm \citep[][]{Branch1999AProblems, Jones2001SciPy:Python} which minimises the sum of squares between the multi-Gaussian model and the observed spectral line profiles. We fit both the He I 5876 and 7065 \mbox{\normalfont\AA} lines simultaneously, and link the velocity of the main absorption component between each of the two lines. By linking the absorption velocities in this way, we constrain the absorption dynamics to represent the information of both lines, leading to a more robust picture of the He I dynamics. These choices are validated by a our multi-Gaussian template producing consistently good fits across all of the spectra. Further checks are made to assess the goodness-of-fit, such as testing whether the normalised residuals show a normal distribution as suggested by \citet[][]{Riener2019GAUSSPY+:Spectra}. Example multi-Gaussian fits are displayed in Figure \ref{fig:HeI7065_multiG_fits}. Here the absorption components, shown in orange, have their velocity linked during the simultaneous fitting of the He I 5876 and 7065 \mbox{\normalfont\AA} lines as described above.

After fitting the line profiles we examine the extracted velocities. For the bulk velocity of \etacar{}'s wind emission region we take the weighted mean of the He I 7065 \mbox{\normalfont\AA} line's four main emission components (centroid wavelength of each Gaussian weighted by its area, see \citet[][]{Blundell2007FluctuationsTimescales, Grant2020Uncovering140}). For the absorption velocities we take the centroid of the velocity-linked absorption component. We plot these velocities in Figure \ref{fig:HeI7065_dynamics} and tabulate them in Table \ref{tab:velocity_table}. Note that the absorption velocities are not present near to periastron as this component is not detected at these times, as was discussed in Section \ref{subsec:he_i_line_profiles}.

In Figure \ref{fig:HeI7065_dynamics} it is clear that the emission and absorption components encode dynamical information from distinct regions of \etacar{}. The emission velocities show similar dynamics to the line-of-sight orbital motion of the system (dashed line), as predicted by the models of \citet[][]{Grant2020Uncovering140, Grant2021ProbabilisticBinaries}. However, the absorption velocities show apparent deviations from the orbital motion. These deviations are most prominent during apastron as the velocities arc upwards to slower values. Understanding these absorption dynamics will now form the focus of the remainder of this study. 

\section{Tracing the Colliding winds}
\label{sec:tracing_the_colliding_winds}
In this section, we formulate a physical model for interpreting the behaviour of the He I absorption and apply this model to the Global Jet Watch observations.

\subsection{The line formation regions of Helium}
\label{subsec:line_formation_regions_of_helium}

In order to model the dynamics of the He I absorption, we must first consider the Helium line formation regions in \etacar{}. The companion is known to be a source of Helium ionising photons \citep[][]{Verner2002TheCarinae, Verner2005TheD, Hillier2006TheCarinae} which significantly affects the Helium ionisation structure of the system; and hence, any quantities deduced from these lines.

\begin{figure}
	\includegraphics[width=\columnwidth]{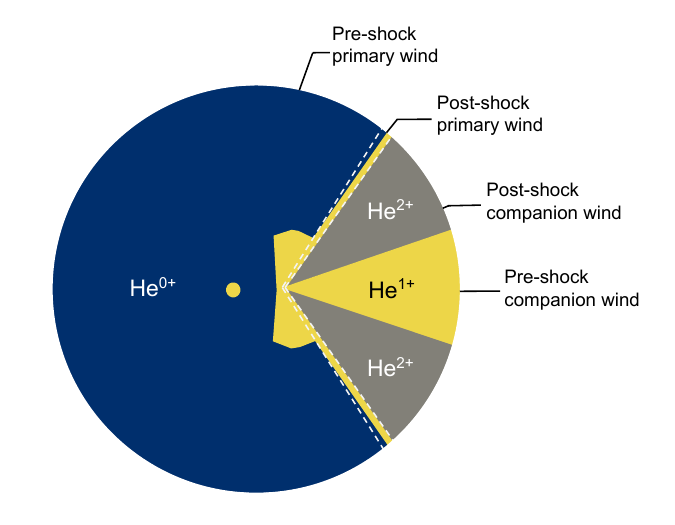}
    \caption{Helium ionisation schematic of \etacar{}: $\rm{He}^{0+}$ (blue), $\rm{He}^{1+}$ (yellow), $\rm{He}^{2+}$ (grey). The radius of the blue region is 155 au and the dashed lines represent the region of the post-shock primary wind formed in the interface of the colliding winds. This plot is reproduced from \citet[][]{Sanchez-Bermudez2018}, which itself was based on the simulations of \citet[][]{Clementel20153DApastron} for apastron.}
    \label{fig:eta_car_he_ionisation_schematic}
\end{figure}

In Figure \ref{fig:eta_car_he_ionisation_schematic} we reproduce the schematic of the Helium ionisation map from \citet[][their figure 14]{Sanchez-Bermudez2018}, which is based on the simulations of \citet[][their figures 7 and 8]{Clementel20153DApastron} for the central 155 au of \etacar{}. In this schematic we colour each region based on its Helium ionisation level: $\rm{He}^{0+}$ (blue), $\rm{He}^{1+}$ (yellow), $\rm{He}^{2+}$ (grey). The primary wind is predominately $\rm{He}^{0+}$, except for two $\rm{He}^{1+}$ regions. The first is the small volume surrounding the primary star which is ionised by radiation directly from the primary star. The second is the region adjacent to the apex of the colliding wind, which is ionised by radiation from the companion. Here we see how only gas sufficiently close to the shock cone is able to be reached by the companion's ionising flux. In addition to these two regions, $\rm{He}^{1+}$ also exists in the post-shock primary wind in the walls of the shock cone cavity, denoted by the dashed lines in Figure \ref{fig:eta_car_he_ionisation_schematic}, and throughout most of the pre-shock companion wind. The post-shock companion wind is solely comprised of $\rm{He}^{2+}$ owing to collisional ionisation caused by the high temperature ($T > 10^6$ K) of the shock-heated gas.

We are primarily concerned with the $\rm{He}^{1+}$ regions of Figure \ref{fig:eta_car_he_ionisation_schematic}, since the observed optical He I lines are indirect consequences of recombination into the highly energetic top levels of these lines. In particular, following \citet[][]{Damineli2008ACarinae}, \citet[][]{Clementel20153DApastron}, and \citet[][]{Sanchez-Bermudez2018}, we presuppose that the He I absorption is formed principally in the post-shock primary wind. 

\subsection{A geometrical model for the colliding winds}
\label{subsec:a_model_for_the_colliding_winds}
There exist certain combinations of system orientations and colliding-wind balances in which the shock cone intersects our line of sight to the primary star. If the He I absorption is formed in the post-shock primary wind, along the walls of the shock cone, then this absorption will trace the geometry of the colliding winds as the shock cone swings across our line of sight. The dynamical information imprinted at the site of the absorption corresponds to the post-shock primary wind velocities projected onto our line of sight. This results in faster velocities the more parallel the shock cone wall is to our line of sight. Qualitatively, for orientations favoured by most investigators, with the shock cone on our side of the system during apastron, this leads to absorption velocities that appear fast as the line of sight first enters the shock cone, shortly after periastron, followed by a progressive slowing as the line of sight traces across to the centre of the shock cone at apastron. The process is then reversed as the line of sight traces out to the other side of the shock cone, the absorption velocities becoming faster, until the line of sight exits the shock cone altogether. This description mirrors the dynamical behaviour of the He I absorption found in Section \ref{subsec:he_i_absorption_dynamics} and so we pursue this line of thinking by formulating a geometric model of the colliding winds.

\begin{figure*}
	\includegraphics[width=\textwidth]{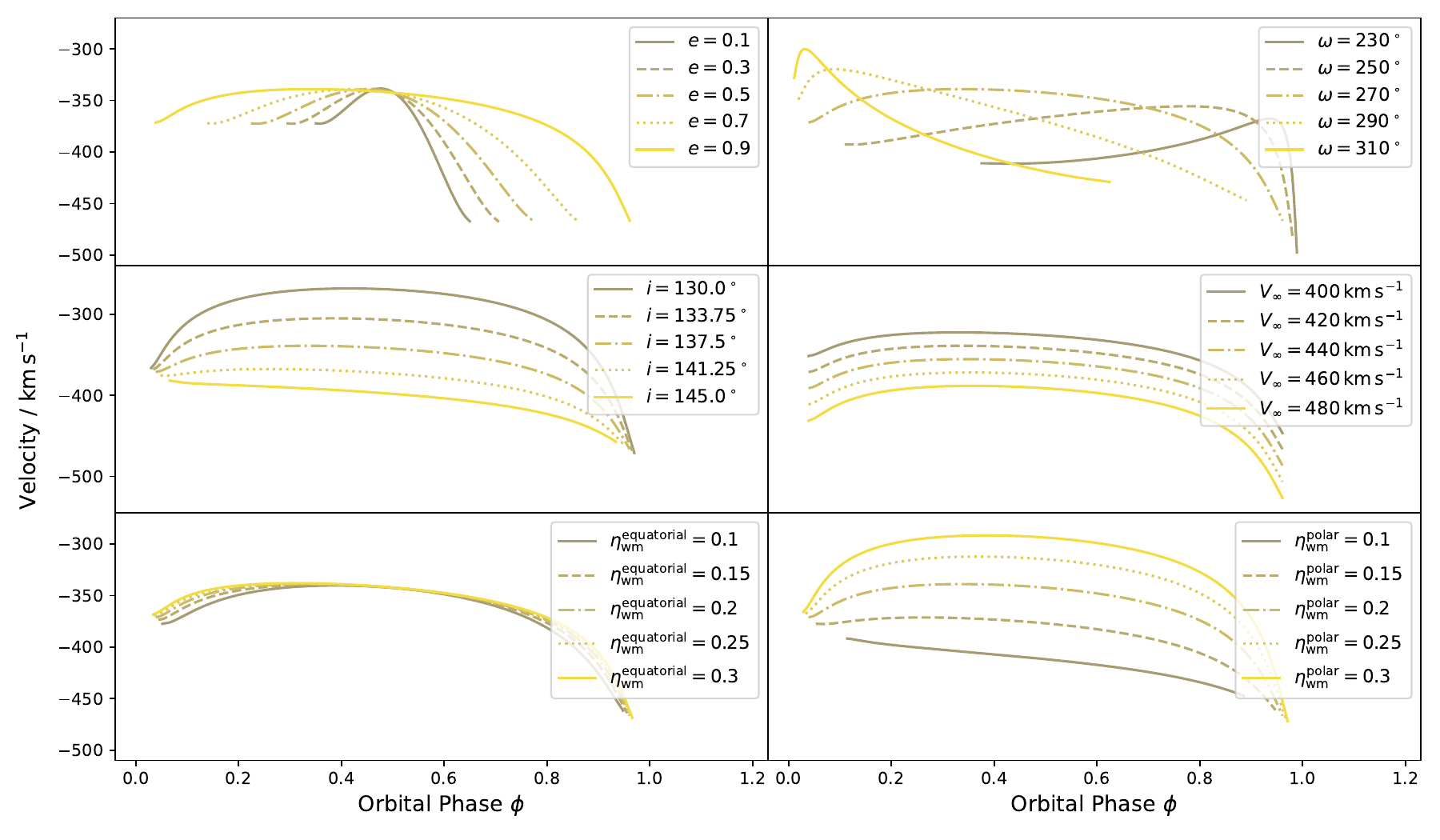}
    \caption{Exploration of our geometrical model for the colliding winds of \etacar{} for various parameter combinations. Only one parameter is varied per panel, see the text for details.}
    \label{fig:colliding_wind_model_exploration}
\end{figure*}

We begin the description of our geometric model by considering the wind momentum balance. Along the line of centres of the two stars the wind momentum ratio is defined as
\begin{equation}
\eta_{\rm{wm}} = \frac{\Dot{M}_{2} v_{\infty, 2}}{\Dot{M}_{1} v_{\infty, 1}},
	\label{eq:tracing_wind_momentum_ratio}
\end{equation}
where $\Dot{M}$ is the mass-loss rate, $v_{\infty}$ is the terminal velocity of the wind, and the subscripts 1 and 2 correspond to the primary and companion stars respectively. Using Equation \ref{eq:tracing_wind_momentum_ratio} the apex of the contact discontinuity occurs at
\begin{equation}
r_1 = \frac{D}{1 + \sqrt{\eta_{\rm{wm}}}},
	\label{eq:tracing_contact_discontinuity_distance}
\end{equation}
where $r_1$ is the distance from the primary star to the contact discontinuity and $D$ is the total separation of the two stars \citep[][]{Stevens1992CollidingSystems}. This expression assumes that the winds collide at their terminal velocities which is valid for most of the apastron passage of \etacar{}. The balance of ram pressures when the stars have uneven wind strengths leads to a cone-like shape for the contact discontinuity, with the asymptotic half-opening angle of this shock cone, $\theta$, being the fundamental descriptor. Previous hydrodynamic simulations by \citet[][]{Madura2013ConstraintsWinds} found that for \etacar{} this $\theta$ was consistent with the analytic formula of \citet[][]{Canto1996ExactProblem}:
\begin{equation}
\eta_{\rm{wm}} = \frac{\tan \theta - \theta}{\tan \theta - \theta + \pi}.
	\label{eq:tracing_contact_discontinuity_angle}
\end{equation}

Given analytical expressions for the colliding wind apex's location and asymptotic opening angle we elect to use a hyperboloid as the geometric surface for the contact discontinuity. A hyperboloid of one sheet, orientated such that the apex and opening are along the $x$-axis, has the following equation:
\begin{equation}
\frac{x^2}{a^2} = \frac{y^2}{b^2} + \frac{x^2}{c^2} + 1,
	\label{eq:hyperboloid}
\end{equation}
where $a = r_1$ sets the distance from the primary star, situated at the origin, to the apex, $b = a \tan \theta$ sets the asymptotic half-opening angle in the $x$-$y$ plane, and $c = a \tan \theta$ sets the asymptotic half-opening angle in the $x$-$z$ plane. In this way we have included the flexibility to alter the wind momentum balance in both the equatorial and polar directions (see the model exploration in Figures \ref{fig:colliding_wind_model_exploration} and \ref{fig:colliding_wind_model_3d_render} and the discussion in Section \ref{sec:tracing_discussion}), although for the main model fitting we set both these values to be the same. A hyperboloid is a natural choice owing to its qualitative agreement with hydrodynamic models, and the ease with which the surface can encode the location and angles of the colliding winds.

To synthesise the absorption velocities of the post-shock primary wind of \etacar{} we work in the rotated frame of reference, which has the line of centres of the two stars fixed along the $x$-axis, and the primary star at the origin. We compute the separation of the stars at each epoch of an orbital period and from this set the hyperboloid geometry after inputting the wind momentum ratio. Our line of sight to the primary star is then computed in this rotated frame and we can solve for the intersection, if any, between it and the hyperboloid, named as point $P_{\rm{int}}$. At $P_{\rm{int}}$ we compute the angle between the tangent plane to the hyperboloid and our line of sight, $\phi$, in order to project the shocked gas velocities onto our line of sight.

The final absorption velocities in the post-shock primary wind, $v_{\rm{absor}}(t)$, as a function of time, $t$, are
\begin{equation}
v_{\rm{absor}}(t) = v_{\rm{wind}}^{\prime}(t, P_{\rm{int}}) \cos^2\phi,
	\label{eq:tracing_absorption_velocities}
\end{equation}
where
\begin{equation}
v_{\rm{wind}}^{\prime}(t, P_{\rm{int}}) = v_{\rm{wind}}(t, P_{\rm{int}}) + v_{\rm{kep}}(t) + \gamma.
	\label{eq:tracing_pre_shock_velocities}
\end{equation}
In these equations $v_{\rm{wind}}$ is the velocity of the primary wind just prior to being shocked at point, $P_{\rm{int}}$, $v_{\rm{kep}}$ is the Keplerian velocity of the primary star projected onto the line-of-sight, and $\gamma$ is the systemic velocity along our line of sight. The sum total of Equation \ref{eq:tracing_pre_shock_velocities} is the velocity of the pre-shock gas in the observer's frame of reference. Equation \ref{eq:tracing_absorption_velocities} simply projects the pre-shock velocities, first along the shock cone, and then second, onto our line of sight. The angles of these projections are both $\phi$ as our line of sight is co-linear with the outflowing pre-shock wind in our line of sight. We also choose to calculate $v_{\rm{wind}}$ from the standard $\beta$-law as first derived by \citet[][]{Castor1975Radiation-drivenStars} for line-driven winds and we set $\beta=1$. However, we find using the terminal velocity of the wind does not affect our results by more than a few $\kmpers$, because at times near to apastron the intersection point is far from the primary star. 

Having described a geometrical model for the colliding winds of \etacar{}, we now look at how the parameters that go into this model change the predicted absorption velocities. The complete parameter set is displayed in Table \ref{tab:tracing_model_parameters}: the orbital period, $P$, time of periastron, $T_0$, eccentricity, $e$, primary star's semi-amplitude, $k_1$, systemic velocity along our line of sight, $\gamma$, wind momentum ratio, $\eta_{\rm{wm}}$, primary wind terminal velocity, $v_{\infty, 1}$, inclination, $i$, and argument of periastron, $\omega$. In addition to these tabulated parameters we also require a value for the mass of the primary star, $m_1$ in order to calculate the size of the semi-major axis via the mass function and Kepler's third law. We set $m_1=100 \solarm$ which results in a semi-major axis of $17.8$ au, but tests show that this choice is arbitrary for our model. The semi-major axis simply rescales the system, having a negligible impact on the absorption velocities, much like the $\beta$-law parameterisation mentioned above. Consequently, we leave this parameter fixed and out of the tabulated list of key parameters in Table \ref{tab:tracing_model_parameters}.

In Figure \ref{fig:colliding_wind_model_exploration} we explore the dependence of the predicted absorption velocities on the model parameters. In each panel we vary one parameter and fix all of the others. The fixed set is $e=0.9$, $k_1 = 66.8 \kmpers$, $\gamma=0 \kmpers$, $v_{\infty, 1} = 420 \kmpers$, $i=137.5^{\circ}$, and $\omega=270^{\circ}$. We further subdivide the wind momentum ratio into its equatorial and polar components, $\eta_{\rm{wm}}^{\rm{equatorial}} = 0.2$ and $\eta_{\rm{wm}}^{\rm{polar}} = 0.2$. We start with the top-left panel where we examine the role of the eccentricity. Here we find the resulting absorption velocities depend strongly on this parameter. The higher the eccentricity, the sooner our line of sight intersects with the shock cone after periastron, and the longer we trace the velocities as the shock cone sweeps more slowly across our line of sight at apastron. In the top-right panel and the middle-left panels we vary the argument of periastron and inclination respectively. Here we find the model is very sensitive to changes in our line of sight. The argument of periastron shifts the orbital phase at which the slowest velocities (largest projection angles) occur. The inclination changes the overall range of velocities predicted, and to a lesser extent the timing of the start and end of the intersection. This can be interpreted as lower inclinations tracing a line through the centre of the shock cone, intersecting it at a variety of angles. However, for higher inclinations the intersection angles remain small throughout, and so the projected velocities do not change much over the orbit. In the middle-right panel we see how the velocity of the primary wind only serves to shift all of the absorption velocities to slower or faster values.

\begin{figure*}
	\includegraphics[width=\textwidth]{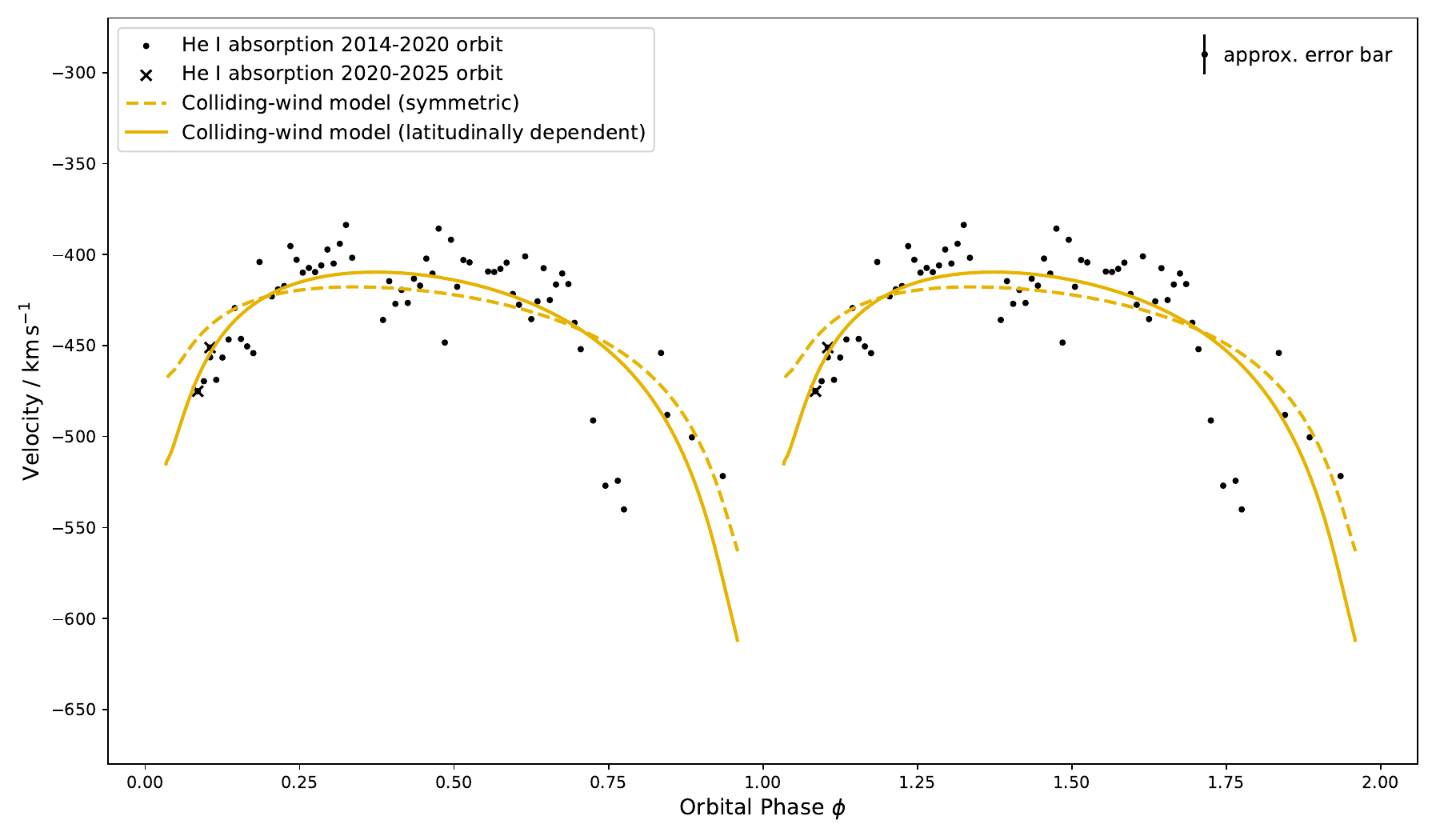}
    \caption{Best fit of our geometrical model for the colliding winds of \etacar{} to the He I absorption velocities. One orbital period has been duplicated. The uncertainties on the velocity data are estimated at $11 \kmpers{}$.}
    \label{fig:colliding_wind_model_best_fit}
\end{figure*}

In the bottom panels of Figure \ref{fig:colliding_wind_model_exploration} we test how the wind momentum ratio affects the predicted absorption velocities of our model. In the bottom-left panel we vary the equatorial ratio and find the model is largely insensitive to this parameter, except for some minor changes to the timing and velocities at the start and end of the intersection. This is the case because, with an inclination set at $137.5^{\circ}$, we trace through the shock cone high above the equatorial plane where the projected angles depend mainly on the polar ratio, as is seen in the bottom-right panel. In the bottom-right panel we find that as the wind momentum ratio increases, meaning the companion star has a more equally balanced wind, the shock cone opening angle increases and the absorption velocities span a larger range. Conversely, for smaller wind momentum ratios our line of sight only just intersects with the cone as it becomes more closed, maintaining faster and more even velocities throughout the orbit. Once the wind momentum ratio becomes sufficiently small there will be no intersection at any time, however this is heavily dependent on the inclination angle as can be seen by comparing their influence on the resulting model in their respective panels. For spherically symmetric outflows we lock the equatorial and polar wind momentum ratios together and the model dependence is the sum of the effects in these two bottom panels. 

\begin{table}
    \centering
    \renewcommand{\arraystretch}{1.3}
    \begin{tabular}{l c c}
    \hline
    Model parameters & Value & Reference \\
    \hline
    \multicolumn{3}{c}{Orbital elements} \\
    \hline
    $P$ [days] & $2022.7$ & \citet[][]{Damineli2008TheEvents} \\
    $T_0$ [JD] & $2454850.1$ & \citet[][]{Grant2021ProbabilisticBinaries} \\
    $e$ & $0.90$ & \citet[][]{Grant2021ProbabilisticBinaries} \\
    $k_1$ $[\kmpers]$ & $66.8$ & \citet[][]{Grant2021ProbabilisticBinaries} \\
    $\gamma$ $[\kmpers]$ & $-8.1$ & \citet[][]{Smith2004a} \\
    \hline
    \multicolumn{3}{c}{Wind parameters} \\
    \hline
    $\eta_{\rm{wm}}$ & $0.2$ & \citet[][]{Pittard2002} \\
    $v_{\infty, 1}$ $[\kmpers]$ & 509 & This work \\
    \hline
    \multicolumn{3}{c}{Line of sight parameters} \\
    \hline
    $i$ [$^{\circ}$] & $137.5$ & \citet[][]{Madura2012ConstrainingEmission} \\
    $\omega$ [$^{\circ}$] & $271$ & This work \\    
    \hline
    \end{tabular}
    \caption{Parameters used in the geometrical model for the colliding winds of \etacar{}. ``This work'' indicates the parameters found in the best-fit symmetric model shown in Figure \ref{fig:colliding_wind_model_best_fit}.}
    \label{tab:tracing_model_parameters}
\end{table}

Next, we fit our model to the He I absorption velocities extracted in Section \ref{subsec:he_i_absorption_dynamics}. We fix the orbital elements to literature values: $P=2022.7$ d \citep[][]{Damineli2008TheEvents}; $T_0=2454850.1$ (JD), $e=0.90$, and $k_1=66.8 \kmpers$ \citep[][]{Grant2021ProbabilisticBinaries}; $\gamma=-8.1 \kmpers$ \citep[][]{Smith2004a}. For the inclination we use a value of $i=137.5^{\circ}$, the central value in the range estimated from models of the [Fe III] emission \citep[][]{Madura2012ConstrainingEmission}, and for the wind momentum ratio  we use a value of $\eta_{\rm{wm}}=0.2$, the value estimated from the duration of the x-ray lightcurve minimum \citep[][]{Pittard2002}. For the model fitting in this section we assume a spherically symmetric wind balance. Finally, parameters $v_{\infty, 1}$ and $\omega$ are left free to be optimised. A summary of this parameterisation is detailed in Table \ref{tab:tracing_model_parameters}.

The model is optimised using the Levenberg-Marquardt algorithm \citep[][]{More1978TheTheory, Jones2001SciPy:Python} to find the least-squares solution. We find the best fit values for our free parameters are $v_{\infty, 1} = 509 \kmpers$ and $\omega = 271^{\circ}$. We display this result in Figure \ref{fig:colliding_wind_model_best_fit}. Additionally, we render the hyperboloid corresponding to this best-fit solution in the top panel of Figure \ref{fig:colliding_wind_model_3d_render}. The model produces good fits to the data, showing the correct velocity trends as well as the correct timing of the entry (absorption first detected after periastron) and exit (absorption no longer detected this orbital cycle) of our line of sight through the shock cone. Although we note that there are several data points at phases $0.72 < \phi < 0.78$ which show faster velocities than those that can be produced by the best-fit model. We tested the model fitting excluding these data and find the results are not sensitive to these four data points.

\begin{table}
    \centering
    \renewcommand{\arraystretch}{1.3}
    \begin{tabular}{l c c}
    \hline
    Model parameters & Prior $\mathcal{N}(\mu)$ & Prior $\mathcal{N}(\sigma)$ \\
    \hline
    \multicolumn{3}{c}{Orbital elements} \\
    \hline
    $e$ & $0.90$ & $0.01$ \\
    $k_1$ $[\kmpers]$ & $66.8$ & $4.25$ \\
    \hline
    \multicolumn{3}{c}{Wind parameters} \\
    \hline
    $\eta_{\rm{wm}}^{\rm{mean}}$ & $\{0.175, 0.20, 0.225, 0.25\}$ & fixed \\
    $\eta_{\rm{wm}}^{\rm{polar}}$ & $0.2$ & $0.05$ \\
    $v_{\infty, 1}$ $[\kmpers]$ & $500$ & $30$ \\
    \hline
    \multicolumn{3}{c}{Line of sight parameters} \\
    \hline
    $i$ [$^{\circ}$] & $\{131, 136, 141, 146\}$ & fixed \\
    $\omega$ [$^{\circ}$] & $270$ & $20$ \\    
    \hline
    \end{tabular}
    \caption{Bayesian priors corresponding to analysis in Section \ref{subsec:tracing_a_latitudinally_dependent_wind}. All parameters listed are given a prior that follows a normal distribution, $\mathcal{N}(\mu, \sigma)$. The inclination priors take on a grid of distributions.}
    \label{tab:tracing_model_bayes_factor_grid}
\end{table}

\begin{figure}
	\includegraphics[width=\columnwidth]{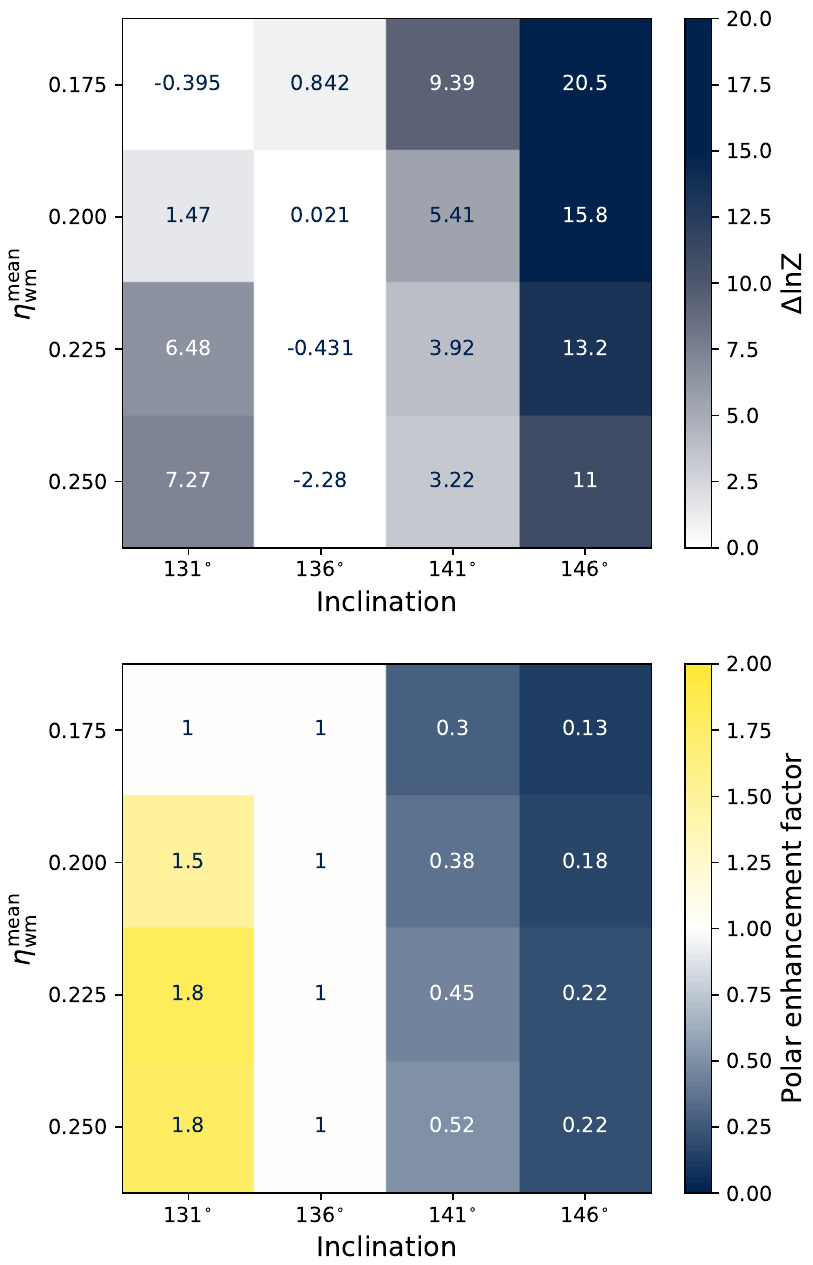}
    \caption{Bayesian analysis grid of model inferences based on varying the inclination and mean wind momentum ratio, $\eta_{\rm{wm}}^{\rm{mean}}$, of the polar and equatorial components. The top panel shows the Bayes factor comparing the evidence for the latitudinally dependent model versus the spherically symmetric model. The bottom panel shows the polar enhancement factor of the wind, defined as $1/(\eta_{\rm{wm}}^{\rm{polar}}/\eta_{\rm{wm}}^{\rm{equatorial}})$, for the statistically preferred model.}
    \label{fig:colliding_wind_model_bayes}
\end{figure}

\section{Discussion}
\label{sec:tracing_discussion}
We have presented a geometrical model for the colliding winds of \etacar{}. The fit of our model to the He I absorption velocities shows good agreement. Moreover, the parameters input into the model are concordant with the other panchromatic observations and modelling of \etacar{}. This includes direct agreement between the Helium ionisation simulations \citep[][]{Clementel20153DApastron}, VLTI/GRAVITY observations of He I 2.0587 $\mu$m, models of the x-ray light curve \citep[][]{Pittard2002}, three-dimensional hydrodynamic simulations \citep[][]{Okazaki2008ModellingCollision}, HST/STIS observations of [Fe III] \citep[][]{Madura2012ConstrainingEmission}, Gemini/GMOS observations of the Balmer lines \citep[][]{Grant2020Uncovering140, Grant2021ProbabilisticBinaries}, and now also the Global Jet Watch observations of He I 5876 and 7065 \mbox{\normalfont\AA}. This prevailing view of \etacar{}'s fundamental parameters builds confidence in our understanding of the system: a highly eccentric colliding-wind binary oriented with the companion on the observer's side of the system during apastron. In the remainder of this section we consider further details of our model within the constraints of this view of \etacar{}.

\subsection{Bayesian evidence of a latitudinally dependent wind}
\label{subsec:tracing_a_latitudinally_dependent_wind}
During the model fitting undertaken in Section \ref{subsec:a_model_for_the_colliding_winds} we assumed spherical symmetry in the colliding winds, locking the wind momentum ratio to identical values in both the $x$-$y$ and $x$-$z$ planes. However, as was shown in the exploration of the model parameters in Figure \ref{fig:colliding_wind_model_exploration}, specifically the middle-left and bottom-right panels, the inclination and wind momentum ratio in the polar direction are severely degenerate. In other words, the projected angles between our line of sight and the shock cone can be held constant by simultaneously decreasing the inclination of the viewing angle and decreasing the polar wind momentum ratio. Consequently, any deductions made about the wind momentum ratio rely on the inclination value, and the inclination for \etacar{} has only been constrained to within a window of $\sim 20^{\circ}$ \citep[][]{Madura2012ConstrainingEmission, Teodoro2016}.

To investigate how this may affect our results we run some further tests. We perform Bayesian inference, using the same geometrical model as used in Section \ref{subsec:a_model_for_the_colliding_winds}, but for a 4-by-4 grid of varying inclination and mean wind momentum ratio. The inclinations span a similar range to those found by \citet[][]{Madura2012ConstrainingEmission} and \citet[][]{Teodoro2016} from $131^{\circ}$ to $146^{\circ}$. The mean wind momentum ratios span a small range centred on the value resulting from analysis of the x-ray light curve \citep[][]{Pittard2002} from 0.175 to 0.250. In this way, we assess how the inferred latitudinal dependence of the wind momentum ratio depends on the inclination constraints.

\begin{figure}
	\includegraphics[trim={1cm 0 1cm 3cm},clip,width=\columnwidth]{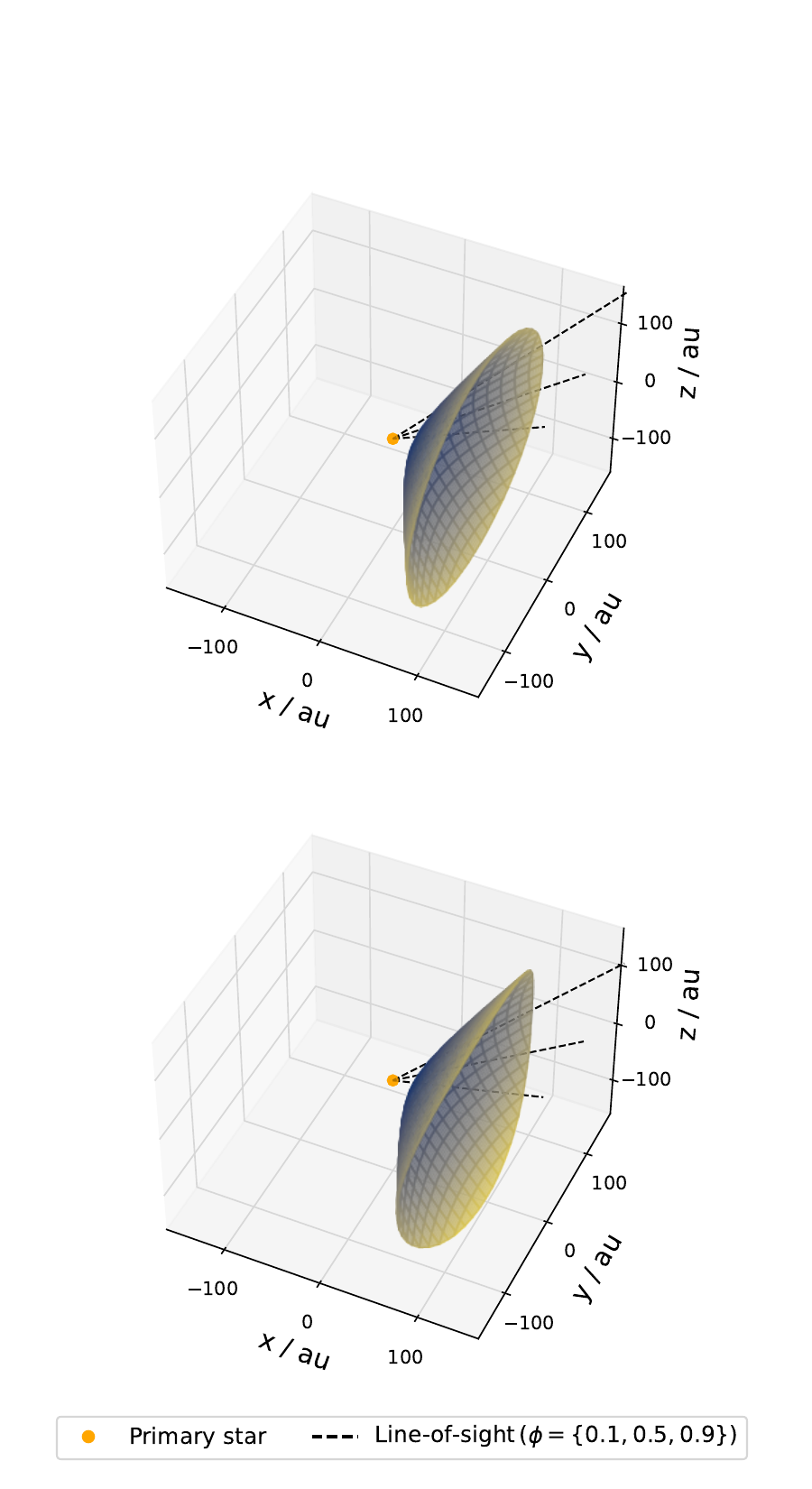}
    \caption{Three-dimensional renderings of the hyperboloid surfaces from our geometric models. The models are shown in the rotated frame of reference with the primary star at the origin. The dashed lines show the line of sight at three different phases: 0.1, 0.5, 0.9. Top panel: the best-fit model from Section \ref{subsec:a_model_for_the_colliding_winds} for a spherically symmetric wind with $i=137.5^{\circ}$ and $\eta_{\rm{wm}} = 0.20$. Bottom panel: an example model from Section \ref{subsec:tracing_a_latitudinally_dependent_wind} for a polar enhanced wind with $i=131^{\circ}$, $\eta_{\rm{wm}}^{\rm{equatorial}} = 0.22$, and $\eta_{\rm{wm}}^{\rm{polar}} = 0.12$.}
    \label{fig:colliding_wind_model_3d_render}
\end{figure}

We run the grid for both the spherically symmetric wind model, but also for a latitudinally dependent wind model, now allowing the polar wind momentum ratio to vary. We parameterise this latitudinally dependent model in terms of $\eta_{\rm{wm}}^{\rm{polar}}$, the polar wind momentum ratio, and $\eta_{\rm{wm}}^{\rm{mean}}$, the mean wind momentum ratio of the polar and equatorial components. The equatorial wind momentum ratio is defined as $\eta_{\rm{wm}}^{\rm{equatorial}} = 2 \eta_{\rm{wm}}^{\rm{mean}} - \eta_{\rm{wm}}^{\rm{polar}}$. This parameterisation ensures the latitudinally dependent model remains consistent with the x-ray light curve observations of \citet[][]{Pittard2002}. A full list of parameters and their priors are given in Table \ref{tab:tracing_model_bayes_factor_grid}.

We use the nested sampling code \texttt{dynesty} \citep[][]{Speagle2020DYNESTY:Evidences} to infer the posterior parameter distributions and model evidence (or marginalised likelihood), $Z$, for both the spherical and latitudinally dependent models at each point in the grid. We use the Bayes factor, $\Delta \ln Z = \ln Z_{\rm{LD}} - \ln Z_{\rm{SS}}$, to select the statistically preferred model, where $Z_{\rm{SS}}$ and $Z_{\rm{LD}}$ are the model evidences for the spherically symmetric and latitudinally dependent models, respectively. For grid points where $\Delta \ln Z > 1.15$, we determine there to be substantial evidence \citep[][]{Kass1995BayesRobert} for the latitudinally dependent model and we infer different wind momentum ratios for the equatorial and polar directions. For $\Delta \ln Z \leq 1.15$, we infer the wind momentum ratio to be spherically symmetric.

The grid of results are presented in Figure \ref{fig:colliding_wind_model_bayes}. The top panel shows the Bayes factor results and we find that many of the grid points do show statistical evidence for a latitudinal wind. In the bottom panel we show the corresponding polar enhancement factor of the wind, defined as $1/(\eta_{\rm{wm}}^{\rm{polar}}/\eta_{\rm{wm}}^{\rm{equatorial}})$. For grid points that have spherically symmetric models preferred, the polar and equatorial wind momentum ratios are equal and the polar enhancement equals one. For those grid points with latitudinally dependent models, we find the wind momentum ratios bifurcate based on the inclination constraints. The polar wind momentum ratio is lower (higher) when the inclination is lower (higher). This means that the primary wind is polar enhanced for inclinations $< 136^{\circ}$ and equatorially enhanced for inclination $>136^{\circ}$, as is shown in Figure \ref{fig:colliding_wind_model_bayes}. Additionally, as the mean wind momentum ratio increases the polar wind becomes stronger relative to the equatorial wind.

The model run with an inclination of $131^{\circ}$ and $\eta_{\rm{wm}}^{\rm{mean}} = 0.20$ is particularly interesting as this scenario leads to $\eta_{\rm{wm}}^{\rm{polar}} = 0.16$ and $\eta_{\rm{wm}}^{\rm{equatorial}} = 0.24$. This implies \etacar{}'s  primary star may have a mass-loss rate of $1.1 \times 10^{-3} \masslossrate$ and $7.1 \times 10^{-4} \masslossrate$ in the polar and equatorial directions, respectively. This stronger polar outflow from the primary star is qualitatively consistent with the observations of the Balmer lines by \citet[][]{Smith2003LatitudedependentCarinae}, as well as theoretical work to include non-radial line-forces and gravity darkening in models of rapidly rotating luminous stars \citep[][]{Cranmer1995TheStars, Owocki1996InhibitionWinds}. We render this case in the bottom panel of Figure \ref{fig:colliding_wind_model_3d_render}. Here we see subtle changes in the hyperboloid surface and line of sight relative to the spherically symmetric case seen in the top panel from Section \ref{subsec:a_model_for_the_colliding_winds}. Although, of course at the opposite side of the inclination grid, for say an inclination of $146^{\circ}$ and $\eta_{\rm{wm}}^{\rm{mean}} = 0.20$, we find the reverse to be true, and a stronger equatorial wind is inferred. Therefore, we highlight the importance of future observations and modelling to improve constraints on the inclination because this will enable us to make quantitative conclusions about the latitudinal dependence of \etacar{}'s winds using the geometrical model presented in this here.

\subsection{Model caveats}
\label{subsec:tracing_model_caveats}
The model and results presented include several assumptions and simplifications worth noting. First, is the omission of the Coriolis effect from the model's geometry. This effect causes the shock cone to trail behind the axis of the two stars; however, it is unlikely to cause noticeable changes to the model at times around apastron as can be seen in hydrodynamic simulations of \etacar{} \citep[][]{Okazaki2008ModellingCollision, Parkin2011SpiralingCarinae, Clementel20153DApastron}. In fact, the instabilities and turbulence in the shocked flows are likely to be a larger source of uncertainty at these times. However, for the few data points at the very start and end of the absorption signal there may be some Coriolis effects, as well as modifications to the Helium ionisation structure \citep[][]{Clementel2015}. This may be manifested in our detection of the Helium absorption only between phases $\phi=0.08$ and $\phi=0.95$, which shows a noticeable asymmetry about periastron, with a delay in our line of sight re-entering the shock cone.

Next, we have simplified the geometry of the colliding winds to the surface of a hyperboloid. Whilst this formalism is certainly a useful tool, and it recovers the asymptotic opening angles at large distances, the exact shape of the apex may differ in reality. We also assumed that the He I absorption only forms in the post-shock primary wind. This is based on the fact that $\rm{He}^{1+}$ is most dense in the post-shock primary wind, more so than in the pre-shock primary wind, and orders of magnitude more so than in the companion's wind which is thought to make no observable contribution to the spectra. This assumption seems to be validated by how well our model fits the data. But there may still be contributions from both the pre-shock and post-shock absorption in the spectra.

\section{Summary and conclusions}
\label{sec:tracing_summary_and_conclusions}
In this study we have investigated the dynamics of the He I 5876 and 7065 \mbox{\normalfont\AA} absorption velocities in the colliding-wind binary \etacar{}. Our work is summarised as follows:
\begin{enumerate}
  \item We made use of Global Jet Watch observations of \etacar{} covering the last 1.3 orbital periods (2630 days). The unprecedented coverage throughout apastron enabled us to extract clear dynamical information at these orbital phases.
  \item We employed a multi-Gaussian fitting algorithm to the He I 5876 and 7065 \mbox{\normalfont\AA} line profiles. We found distinct absorption components depending on the orbital phase. The slower of these two absorption components is only detected between phases $\phi=0.08$ and $\phi=0.95$ ($\phi=0$ is periastron), and displays velocities that deviate from the orbital-like motion which is found in the emission components of this line.
  \item To interpret these deviations we conjectured that this absorption component of the He I  lines is formed in the post-shock primary wind and is only detected when our line of sight intersects with the shock cone. We formulated a geometrical model for the colliding winds in terms of a hyperboloid in which the opening angle and location of its apex are parameterised in terms of the system's wind momentum ratio. The absorption velocities of the post-shock primary wind are computed as the sum of the shocked wind velocities projected onto our line of sight, the orbital motion, and the systemic velocity of the system.
  \item We fitted our geometrical model to the He I line absorption velocities finding results that are concordant with the panchromatic observations and simulations of \etacar{}.
\end{enumerate}
The model presented in this study is an extremely sensitive probe of the exact geometry of the wind momentum balance of binary stars. Given more certainty in the inclination of \etacar{} with respect to our line of sight, the model could be used to probe the latitudinal dependence of the mass loss quantitatively.

\section*{Acknowledgements}
\label{sec:acknowledgements}
We thank the anonymous referee for a helpful and constructive report. A great many organisations and individuals have contributed to the success of the Global Jet Watch observatories and these are listed on \url{http://www.GlobalJetWatch.net} but we particularly thank the University of Oxford and the Australian Astronomical Observatory. We would like to thank Jonathan Patterson for his support of the Oxford Physics computing cluster and the Global Jet Watch servers. We gratefully acknowledge the use of the following software: \texttt{numpy} \citep[][]{VanDerWalt2011TheComputation}, \texttt{pandas} \citep[][]{McKinney2010DataPython}, \texttt{scipy} \citep[][]{Jones2001SciPy:Python}, \texttt{barycorrpy} \citep[][]{Kanodia2018Pythonbarycorrpy}, \texttt{dynesty} \citep[][]{Speagle2020DYNESTY:Evidences}, and \texttt{matplotlib} \citep[][]{Hunter2007Matplotlib:Environment}.

\section*{Data availability}
\label{sec:data_availability}
This research has made use of data from the Global Jet Watch. The data underlying this article will be shared on reasonable request to K. Blundell.

%%%%%%%%%%%%%%%%%%%%%%%%%%%%%%%%%%%%%%%%%%%%%%%%%%



\bibliographystyle{mnras}
\bibliography{references} % if your bibtex file is called example.bib

%%%%%%%%%%%%%%%%%%%%%%%%%%%%%%%%%%%%%%%%%%%%%%%%%%

%%%%%%%%%%%%%%%%% APPENDICES %%%%%%%%%%%%%%%%%%%%%

\appendix

\begin{table*}
    \centering
    \renewcommand{\arraystretch}{0.80}
    \begin{tabular}{c c c c c c}
    \hline
    JD & Phase & Emission velocity & Emission $\sigma$ & Absorption velocity & Absorption $\sigma$ \\
    \hline
2456803.9372771992 & 0.96 & -137.0 & 8.4 & $\cdots$ & $\cdots$ \\
2456829.9486315884 & 0.97 & -154.0 & 8.4 & $\cdots$ & $\cdots$ \\
2456836.9816454477 & 0.98 & -171.2 & 8.4 & $\cdots$ & $\cdots$ \\
2457050.7208976992 & 0.08 & -49.5 & 8.4 & $\cdots$ & $\cdots$ \\
2457063.9381872457 & 0.09 & -38.9 & 8.4 & -469.6 & 11.2 \\
2457088.9534360226 & 0.10 & -55.0 & 8.4 & -456.4 & 11.2 \\
2457102.8352715042 & 0.11 & -68.8 & 8.4 & -468.8 & 11.2 \\
2457120.8128659185 & 0.12 & -65.6 & 8.4 & -456.5 & 11.2 \\
2457144.8505931710 & 0.13 & -63.7 & 8.4 & -446.6 & 11.2 \\
2457166.9855833338 & 0.14 & -61.8 & 8.4 & -429.3 & 11.2 \\
2457186.9173628720 & 0.15 & -63.9 & 8.4 & -446.3 & 11.2 \\
2457207.3406912140 & 0.16 & -68.0 & 8.4 & -450.4 & 11.2 \\
2457222.8659953710 & 0.17 & -74.3 & 8.4 & -454.1 & 11.2 \\
2457241.3295023150 & 0.18 & -69.7 & 8.4 & -404.0 & 11.2 \\
2457289.7493097020 & 0.20 & -77.5 & 8.4 & -422.9 & 11.2 \\
2457302.7928365385 & 0.21 & -82.2 & 8.4 & -419.1 & 11.2 \\
2457324.7436517957 & 0.22 & -68.6 & 8.4 & -417.3 & 11.2 \\
2457342.3517303240 & 0.23 & -78.3 & 8.4 & -395.3 & 11.2 \\
2457373.8166263130 & 0.24 & -84.3 & 8.4 & -402.8 & 11.2 \\
2457381.1167290660 & 0.25 & -79.9 & 8.4 & -409.9 & 11.2 \\
2457406.8799037700 & 0.26 & -79.2 & 8.4 & -407.3 & 11.2 \\
2457430.2075498510 & 0.27 & -86.4 & 8.4 & -409.6 & 11.2 \\
2457445.4796396894 & 0.28 & -81.2 & 8.4 & -405.9 & 11.2 \\
2457466.0579699078 & 0.29 & -73.9 & 8.4 & -397.2 & 11.2 \\
2457487.6064215070 & 0.30 & -86.7 & 8.4 & -404.9 & 11.2 \\
2457512.3197004627 & 0.31 & -84.1 & 8.4 & -394.0 & 11.2 \\
2457539.0980138890 & 0.32 & -78.4 & 8.4 & -383.7 & 11.2 \\
2457541.5593392253 & 0.33 & -86.3 & 8.4 & -401.7 & 11.2 \\
2457658.9104156606 & 0.38 & -89.2 & 8.4 & -435.9 & 11.2 \\
2457664.5975942463 & 0.39 & -75.0 & 8.4 & -414.6 & 11.2 \\
2457691.1298716934 & 0.40 & -76.1 & 8.4 & -427.0 & 11.2 \\
2457709.6832274306 & 0.41 & -76.1 & 8.4 & -419.4 & 11.2 \\
2457727.1710474540 & 0.42 & -107.4 & 8.4 & -426.5 & 11.2 \\
2457758.0133771930 & 0.43 & -86.6 & 8.4 & -413.2 & 11.2 \\
2457774.6663876030 & 0.44 & -79.6 & 8.4 & -417.0 & 11.2 \\
2457793.2952323080 & 0.45 & -86.8 & 8.4 & -402.2 & 11.2 \\
2457813.6994308745 & 0.46 & -86.5 & 8.4 & -410.4 & 11.2 \\
2457835.7748437500 & 0.47 & -107.4 & 8.4 & -385.7 & 11.2 \\
2457849.2080343366 & 0.48 & -91.9 & 8.4 & -448.3 & 11.2 \\
2457874.3125231476 & 0.49 & -100.2 & 8.4 & -391.8 & 11.2 \\
2457894.2076317663 & 0.50 & -94.7 & 8.4 & -417.6 & 11.2 \\
2457912.4237699560 & 0.51 & -94.9 & 8.4 & -402.9 & 11.2 \\
2457933.4959053495 & 0.52 & -90.6 & 8.4 & -404.2 & 11.2 \\
2457998.8875405090 & 0.55 & -83.2 & 8.4 & -409.3 & 11.2 \\
2458011.8980810186 & 0.56 & -92.9 & 8.4 & -409.5 & 11.2 \\
2458031.1978240740 & 0.57 & -85.7 & 8.4 & -407.8 & 11.2 \\
2458054.9110364780 & 0.58 & -91.6 & 8.4 & -404.4 & 11.2 \\
2458068.7483294750 & 0.59 & -109.2 & 8.4 & -421.6 & 11.2 \\
2458095.4551076390 & 0.60 & -111.5 & 8.4 & -427.5 & 11.2 \\
2458116.9557581020 & 0.61 & -73.7 & 8.4 & -400.9 & 11.2 \\
2458140.0500907250 & 0.62 & -93.1 & 8.4 & -435.4 & 11.2 \\
2458156.0781986530 & 0.63 & -105.7 & 8.4 & -425.7 & 11.2 \\
2458170.3133680555 & 0.64 & -104.4 & 8.4 & -407.4 & 11.2 \\
2458198.0009006737 & 0.65 & -122.2 & 8.4 & -425.0 & 11.2 \\
2458216.7155031680 & 0.66 & -104.2 & 8.4 & -416.4 & 11.2 \\
2458243.5539765214 & 0.67 & -105.3 & 8.4 & -410.3 & 11.2 \\
2458255.9642063490 & 0.68 & -111.5 & 8.4 & -416.2 & 11.2 \\
2458280.6079606480 & 0.69 & -96.6 & 8.4 & -437.5 & 11.2 \\
2458292.0064814817 & 0.70 & -118.5 & 8.4 & -452.0 & 11.2 \\
2458334.8816493056 & 0.72 & -107.9 & 8.4 & -491.1 & 11.2 \\
2458385.5618634260 & 0.74 & -101.0 & 8.4 & -527.0 & 11.2 \\
2458442.4152772636 & 0.77 & -111.1 & 8.4 & -540.0 & 11.2 \\
2458566.8734179353 & 0.83 & -124.1 & 8.4 & -454.0 & 11.2 \\
2458578.7888917400 & 0.84 & -123.8 & 8.4 & -488.1 & 11.2 \\
2458669.5510539496 & 0.88 & -119.6 & 8.4 & -500.4 & 11.2 \\
2458765.8721298000 & 0.93 & -139.7 & 8.4 & -521.7 & 11.2 \\
2458789.4541054060 & 0.94 & -153.4 & 8.4 & $\cdots$ & $\cdots$ \\
2458803.7271773710 & 0.95 & -128.1 & 8.4 & $\cdots$ & $\cdots$ \\
2458826.7498016820 & 0.96 & -138.7 & 8.4 & $\cdots$ & $\cdots$ \\
2458841.5475941500 & 0.97 & -153.6 & 8.4 & $\cdots$ & $\cdots$ \\
2458922.1927467366 & 0.01 & -82.7 & 8.4 & $\cdots$ & $\cdots$ \\
2458948.9432226424 & 0.02 & -59.2 & 8.4 & $\cdots$ & $\cdots$ \\
2458962.6176150455 & 0.03 & -46.9 & 8.4 & $\cdots$ & $\cdots$ \\
2458979.6881390084 & 0.04 & -35.5 & 8.4 & $\cdots$ & $\cdots$ \\
2459004.5194909214 & 0.05 & -45.2 & 8.4 & $\cdots$ & $\cdots$ \\
2459031.2815254130 & 0.06 & -82.1 & 8.4 & $\cdots$ & $\cdots$ \\
2459047.9281711860 & 0.07 & -62.5 & 8.4 & $\cdots$ & $\cdots$ \\
2459060.5376633040 & 0.08 & -65.5 & 8.4 & -475.0 & 11.2 \\
2459084.7622875280 & 0.09 & -59.6 & 8.4 & $\cdots$ & $\cdots$ \\
2459109.9452039570 & 0.10 & -42.1 & 8.4 & -451.1 & 11.2 \\
    \hline
    \end{tabular}
    \caption{He I velocities as described in Section \ref{subsec:he_i_absorption_dynamics}. Velocity units are $\kmpers$. Errors are estimated from the mean covariance matrix output from the multi-Gaussian fitting.}
    \label{tab:velocity_table}
\end{table*}

%%%%%%%%%%%%%%%%%%%%%%%%%%%%%%%%%%%%%%%%%%%%%%%%%%

% Don't change these lines
\bsp	% typesetting comment
\label{lastpage}
\end{document}